\documentclass[twocolumn,floats,prl,aps,showpacs]{revtex4}
\usepackage{graphicx}
\usepackage{amsfonts}
\usepackage{amsmath}
\usepackage{amssymb}
\usepackage{exscale}
\usepackage{color}

\def\lan{\langle}
\def\ran{\rangle}

\def\dag{\dagger}

\def\vQ{{\bf Q}}

\newcommand{\bd}{\begin{equation}}
\newcommand{\ed}{\end{equation}}
\newcommand{\be}{\begin{equation}}
\newcommand{\ee}{\end{equation}}
\newcommand{\bt}{\begin{split}}
\newcommand{\et}{\end{split}}

\newcommand{\bn}{\begin{align}}
\newcommand{\en}{\end{align}}
\newcommand{\bea}{\begin{eqnarray}}
\newcommand{\eea}{\end{eqnarray}}
\newcommand{\ba}{\begin{array}}
\newcommand{\ea}{\end{array}}
\newcommand{\nn}{\nonumber}

\begin{document}

\title{Coboson many-body approach to the $N$-exciton ground-state energy}
\author{Shiue-Yuan Shiau$^1$, Yia-Chung Chang$^{2,1}$, Monique Combescot$^3$}

\affiliation{(1) Department of Physics and National Center of Theoretical Sciences, National Cheng Kung University, Tainan, 701 Taiwan}
\affiliation{(2)Research Center for Applied Sciences, Academia Sinica, Taipei, 115 Taiwan}
\affiliation{(3)Institut des NanoSciences de Paris, Universit\'e Pierre et Marie Curie,
CNRS, Tour 22, 2 place Jussieu, 75005 Paris}
\date{\today }


\begin{abstract}
We derive the ground-state energy of $N$ composite bosons made of fermion pairs using the recently developed composite boson many-body formalism. We concentrate on the $N$-pair energy linear in density. We show that the scattering relevant for scattering length contains not only direct and exchange interaction scatterings but also the dimensionless ``Pauli scattering" for fermion exchange in the absence of fermion-fermion interaction. Numerical resolution of the resulting ``ladder" integral equation for fermions interacting through long-range Coulomb forces --- which act as effective repulsion between excitons made of same-spin electrons and same-spin holes --- shows that the prefactor of the $N$-exciton energy linear in density is substantially decreased from its Born approximation value, $13\pi/3$, by a factor $\simeq0.38$ for equal electron and hole effective masses. Interestingly, this factor goes to zero when the hole mass goes to infinity, making the triplet-exciton gas unstable in
 this limit.

\end{abstract}

\pacs{03.75.Hh}

\maketitle


  Although paired fermions can be approached through elementary fermion many-body formalisms developed in the 50's, these formalisms are not suitable to handle fermion exchange between composite bosons made of fermion pairs.\

Various ``bosonization" procedures have been proposed to transform composite bosons into elementary bosons interacting through effective scatterings in which a certain amount of fermion exchange is included\cite{Klein1991}. This can produce correct result to some problems but fails to address all of them, even in the dilute limit, for a very simple dimensional argument: while scatterings appearing in effective Hamiltonians are \textit{energy-like} quantity by construction, the many-body physics of paired fermions is driven by \textit{dimensionless} ``Pauli scatterings"  which originate from the Pauli exclusion principle in the absence of fermion-fermion interaction.\

  A decade and a half ago, we have proposed a conceptually new many-body formalism which treats composite bosons (``cobosons") as entity while keeping exchange between their fermionic components\cite{MoniqPhysreport}. This coboson formalism relies on an operator algebra instead of Green function scalars. It has been mainly used to understand and better predict non-linear optical effects in semiconductors\cite{MoniqEPL2005}, these effects being controlled by fermion exchange in the absence of Coulomb process.\

  In this Letter, we use this coboson many-body formalism to address the ground-state energy of $N$ composite bosons made of fermion pairs, more precisely excitons made of same-spin electrons and same-spin holes.\

  It is known that the ground-state energy of $N$ interacting elementary bosons has a linear term in density which comes from the repeated interaction of two out of $N$ bosons. The next term in density comes from singular correlation\cite{Fetter,Lee1957,Brueckner1957} between three and more bosons through ``bubble" processes associated with same-momentum transfer excitations from the condensate\cite{Sean}. In the case of composite bosons, we expect the energy linear in density to also come from repeated interaction between two out of $N$ cobosons. The challenge is to determine how fermion exchange enters the associated scattering.

  \begin{figure}[t!]
\begin{center}
   \includegraphics[trim=2cm 4.2cm 1.5cm 4cm,clip,width=3.4in] {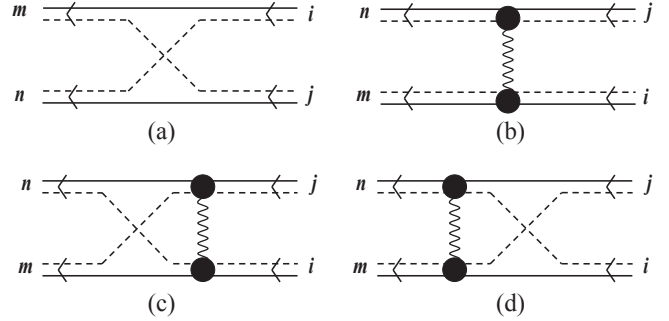}
   \caption{\small (a) Pauli scattering $\lambda(_{mi}^{\, nj})$. (b) Direct Coulomb scattering $\xi(_{mi}^{\, nj})$. (c) ``In" exchange Coulomb scattering $\xi^{in}(_{mi}^{\, nj})$. (d) ``Out" exchange Coulomb scattering $\xi^{out}(_{mi}^{\, nj})$. }
   \label{Fig1}
\end{center}
\end{figure}

  The coboson many-body formalism generates two conceptually different scatterings between $(i,j)$ and $(m,n)$ states: (i) the \textit{dimensionless} ``Pauli scattering" $\lambda(_{mi}^{\, nj})$, shown in Fig.~\ref{Fig1}(a), when the state change results from fermion exchange in the absence of fermion-fermion interaction; (ii) the \textit{energy-like} direct interaction scattering $\xi(_{mi}^{\, nj})$, shown in Fig.~\ref{Fig1}(b), when the state change results from fermion-fermion interaction in the absence of fermion exchange --- fermions being paired in the same way in $(i,j)$ and $(m,n)$ states. All many-body effects involving cobosons can be derived in terms of these two elementary scatterings.

   We here show that the scattering between $(i,j)$ and $(m,n)$ coboson pairs appearing in the ``ladder" processes leading to the $N$-coboson energy linear in density reads as
   \begin{equation}
            \zeta(_{mi}^{\, nj})
     =\xi(_{mi}^{\, nj})-\xi^{exch}(_{mi}^{\, nj})-\frac{E_{mn}+E_{ij}-2E_{00}}{2}\lambda(_{mi}^{\, nj})\, ,\label{eq:zeta}
        \end{equation}
        $E_{ij}-E_{00}=E_i+E_j-2E_0$ is the excitation energy of the coboson pair $(i,j)$ outside the condensate, $i=0$ denoting the single coboson ground state. $\xi^{exch}(_{mi}^{\, nj})$ is the average exchange interaction scattering
        \begin{equation}
            \xi^{exch}(_{mi}^{\, nj})=\frac{ \xi^{in}(_{mi}^{\, nj})+ \xi^{out}(_{mi}^{\, nj})}{2} \, .
        \end{equation}
         In $\xi^{in}(_{mi}^{\, nj})$, interaction takes place between the ``in" states $(i,j)$
        \begin{equation}
            \xi^{in}(_{mi}^{\, nj})
     =\sum_{pq} \lambda(_{mp}^{\, nq}) \xi(_{q\,i}^{\, pj})
        \end{equation}
   (see Fig.~\ref{Fig1}(c)) while in $\xi^{out}(_{mi}^{\, nj})$, interaction takes place between the ``out" states $(m,n)$; so, $\xi^{out}$ reads as $\xi^{in}$  with $\xi$ and $\lambda$ interchanged (see Fig.~\ref{Fig1}(d)).

        For readers knowledgeable with coboson many-body formalism, the scattering between $(i,j)$ and $(m,n)$ states relevant in the $N$-coboson energy cannot have a form different from Eq.~(\ref{eq:zeta}). Indeed, this scattering appears in matrix elements involving the fermionic Hamiltonian $H$ between coboson states. For $H$ acting on the right, the direct scattering $\xi$ leads to $-\xi^{in}$ due to fermion exchange when scalar products of coboson states are taken. By contrast, for $H$ acting on the left, $\xi$ leads to $-\xi^{out}$. As these two exchange interaction scatterings are related through Pauli scattering as\cite{MoniqPhysreport}
        \begin{equation}
            \xi^{in}(_{mi}^{\, nj})-\xi^{out}(_{mi}^{\, nj})
     =\Big(E_{mn}-E_{ij}\Big)\lambda(_{mi}^{\, nj})\, .\label{rel:xiinxiout}
        \end{equation}
        $E\, \lambda$ terms must appear along with $\xi^{in}$ or  $\xi^{out}$ in order to get the same result for $H$ acting on the right or on the left. Precise calculations show that they do.  Then, the scattering given in Eq.~(\ref{eq:zeta}) is the one having the required time reversal symmetry
        \begin{equation}
            \zeta(_{mi}^{\, nj})
     =\Big(\zeta(_{im}^{\, jn})\Big)^*\, .
     \end{equation}

     In the case of excitons, the effective scattering\cite{HH} produced by bosonization, namely $\xi(_{mi}^{\, nj})-\xi^{in}(_{mi}^{\, nj})$ in our notations, should have been rejected long ago because it induces a (missed) non-hermiticity\cite{MCOBM} in the Hamiltonian as $(\xi^{in}(_{mi}^{\, nj}))^*=\xi^{out}(_{im}^{\, jn})$. Hermiticity would be easy to restore with $\xi^{in}$ replaced by $\xi^{exch}$; but the $E\lambda$ terms of $\zeta$ are less obvious. This once more shows the weakness of bosonization in problems dealing with coboson many-body effects.\

     The part of $N$-coboson energy resulting from ladder processes between two cobosons among $N$ can be written in terms of the $\zeta$ scattering as
     \be
\mathcal{E}_N-NE_{0}\simeq \frac{N(N-1)}{2}\bigg(\zeta(_{0\,0}^{0\,0})+\sum_{ij\neq 00}\frac{ \zeta(_{0\,i}^{0\,j})\hat \zeta(_{i\,0}^{j\,0})
}{E_{00}-E_{ij}}\bigg)\,  \label{eq:mathEN-NE0}
\ee
with $\hat \zeta$ solution of the ``ladder" integral equation
\be
\hat \zeta(_{m\,0}^{n\,0})= \zeta(_{m\,0}^{n\,0})+\sum_{ij\neq 00}\frac{ \zeta(_{m\,i}^{n\,j})\hat \zeta(_{i\,0}^{j\,0})}{E_{00}-E_{ij}}\, .\label{eq:zetaij00}
\ee
These two equations, along with Eq.~(\ref{eq:zeta}), constitute the key results of the Letter.



 Just as the $N(N-1)$ prefactor in Eq.~(\ref{eq:mathEN-NE0}) comes from the number of ways to choose the two cobosons which interact among $N$, processes involving three cobosons
 will appear with a $N(N-1)(N-2)$ prefactor; and so on $\ldots$ Summing up these higher-order terms produces the $N$-coboson correlation energy, which is expected to be singular as the one for elementary bosons. Its study is beyond the scope of the present work.

We wish to note that, as $N(N-1)/2=1$ for $N=2$ while Eq.~(\ref{eq:mathEN-NE0}) is valid for arbitrary $N$,
 the part of $N$-pair energy linear in density is related to the $2$-pair energy through
    \be
\mathcal{E}_N-NE_{0}\simeq \frac{N(N-1)}{2}\Big(\mathcal{E}_2-2E_{0}\Big)\, .
\ee
So, to get $\mathcal{E}_N$ at first order in density, or equivalently the coboson scattering length, we can either solve the integral equation (\ref{eq:zetaij00}) or compute the $2$-pair ground state energy by solving the corresponding $4$-fermion Schr\"odinger equation --- as previously done for short-range potential in the cold quantum gas context\cite{Petrov2004,Leyronas2007,Alzetto2013}.

  Let us now outline how the above results follow from the coboson many-body formalism. We look for the ground state of $N$ composite bosons
\be
\big(H-\mathcal{E}_N\big)|\Psi_N\ran=0\, .\label{eq:Scrhodingertwoexciton}
\ee
In the dilute limit, we expect $|\Psi_N\ran\simeq|\Phi_N\ran=B^{\dag N}_0|v\ran$ within states having some cobosons outside the condensate ($|v\ran$ denotes the vacuum state). This leads us to look for $|\Psi_N\ran$ as
\be
|\Psi_N\ran=|\Phi_N\ran+\sum_{ij\neq 00}c_{ij}B^\dag_iB^\dag_j|\Phi_{N-2}\ran+\cdots\label{psiNexpansion}
\ee
where $B^\dag_i$ creates a single coboson, $(H-E_i)B^\dag_i|v\ran=0$. To go further, we use two key commutators of the coboson many-body formalism, namely\cite{MoniqPhysreport}
\bea
\big[H,B^\dag_i\big]_-&=&E_iB^{\dag}_{i}+ V^{\dag}_{i}\, ,\label{eq:HB}\\
\big[
V^{\dag}_{i},B^{\dag}_{j}
\big]_-
&=&\sum_{mn} B^{\dag}_{m}B^{\dag}_{n}\xi (_{mi}^{\, nj})\, ,\label{eq:VB}
      \eea
 the direct interaction scattering  $\xi (_{mi}^{\, nj})$ being possibly replaced by $- \xi^{in}(_{mi}^{\, nj})$
        or even by
 \be
 \Xi(_{mi}^{\, nj})=\frac{a\xi(_{mi}^{\, nj})-b\xi^{in}(_{mi}^{\, nj})}{a+b}\, .
  \end{equation}

The $(a,b)$ indetermination at this level of calculation is profound. It comes from the fact that, due to fermion indistinguishability, there is no way to know of which fermion pair a composite boson is made. This unpleasant feature is omnipresent in the coboson many-body formalism because fermion exchange allows transforming a pair of coboson creation operators $B^{\dag}_{i}B^{\dag}_{j}$ into a sum of other coboson operators according to\cite{MoniqPhysreport}
 \begin{equation}
B^{\dag}_{i}B^{\dag}_{j}=-\sum_{mn} B^{\dag}_{m}B^{\dag}_{n}\lambda(_{mi}^{\, nj})\, .\label{BBfermionexchage}
      \end{equation}
    The above equation, which manifests two possible ways to form two cobosons out of two fermion pairs, readily changes $\xi (_{mi}^{\, nj})$  into $- \xi^{in}(_{mi}^{\, nj})$ in Eq.~(\ref{eq:VB}). Of course, physical results do not depend on $(a,b)$.

    To get $H|\Psi_N\ran$ in Eq.~(\ref{eq:Scrhodingertwoexciton}), we iterate Eqs.~(\ref{eq:HB},\ref{eq:VB}). This yields
    \be
\Big(H-NE_0\Big)|\Phi_N\ran=\frac{N(N-1)}{2}\sum_{mn}B^\dag_mB^\dag_n |\Phi_{N-2}\ran \,
\Xi(_{m0}^{\,n0})\, .\label{eq:H-NE0=xi}
\ee
and, for $(i,j)\neq (0,0)$
 \bea
\lefteqn{\Big(H-E_{ij}-(N-2)E_0\Big) B^\dag_iB^\dag_j |\Phi_{N-2}\ran}\hspace{7.5cm}\nn\\
\simeq \sum_{pq}B^\dag_pB^\dag_q |\Phi_{N-2}\ran \,
\Xi(_{pi}^{qj})\, , \label{eq:H-EijN-2}
 \eea
 within terms in $(N-2)$ and higher which correspond to processes involving more than two cobosons excited from the condensate. These higher-order processes lead to correlation energy, as previously explained. So, we shall not consider them.\

 Next, we project the Schr\"{o}dinger equation (\ref{eq:Scrhodingertwoexciton}) onto $|\Phi_N\ran$ and
 $B^\dag_m B^\dag_n |\Phi_{N-2}\ran$. These projections make appear scalar products of coboson states. Starting from $\lan \Phi_N|\Phi_N\ran\ =N!F_N$ [see Ref.~\onlinecite{MoniqPhysreport}], we get
\be
\lan \Phi_{N-2}|B_n B_m|\Phi_N\ran\simeq  N!F_{N-2}\Big(\delta_{m0}\delta_{n0}-\lambda(_{m0}^{\, n0})\Big)\, ,\label{BnBmphiN}
\ee
and
\bea
\lefteqn{\lan \Phi_{N-2}|B_nB_mB^\dag_i B^\dag_j |\Phi_{N-2}\ran}\hspace{7.7cm}\nn\\
\simeq  (N-2)!F_{N-2}\Big(\delta_{mi}\delta_{nj}-\lambda(_{mi}^{\, nj})+(i\longleftrightarrow j)\Big)\, ,\label{eq:BBBB}
\eea
within terms in $(N-2)$. These scalar products follow from the iteration of the other two key commutators of the coboson many-body formalism, namely
\bea
\big[B_m,B^{\dag}_{j}\big]_-&=&\delta_{mi}-D_{mi}\, ,\\
\big[D_{mi},B^{\dag}_{j}\big]_-&=&\sum_nB^\dag_n \Big(
\lambda(_{mi}^{\, nj})+(i\longleftrightarrow j) \Big)\, .
\eea

 The projection of Eq.~(\ref{eq:Scrhodingertwoexciton}) over $|\Phi_N\ran$, for $|\Psi_N\ran$ written as in Eq.~(\ref{psiNexpansion}), then reads
\be
0=\lan \Phi_N|H-\mathcal{E}_N|\Phi_N\ran +\sum_{ij\neq 00}T(_{0i}^{0j})c_{ij}\,
\label{eq:H-NERN}
\ee
with $T(_{0i}^{0j})$ given by
\bea
T(_{0i}^{0j})=\Bigg\{ N!F_{N-2}\Big(\delta_{0i} \delta_{0j}-\lambda(_{0i}^{ 0j}) \Big) \Bigg\}
 \Big(E_{ij}-E_{00}-\Delta_N\Big)
 \nn\\
\,\,\,\,\,+\sum_{pq}\Bigg\{ N! F_{N-2} \Big(\delta_{0p}\delta_{0q}-\lambda(_{0p}^{ 0q}) \Big)\Bigg\} \Xi(_{pi}^{\,qj})\,\,\,\,\,\,\,\,\,\,\,\,\,\,\,
\eea
for $\Delta_N{=}\mathcal{E}_N{-}NE_0$. To go further, we note that
\be
\sum_{pq}\Big(\delta_{mp}\delta_{nq}-\lambda(_{mp}^{\, nq})\Big) \Xi(_{pi}^{qj})= \xi(_{mi}^{\,nj})- \xi^{in}(_{mi}^{\,nj})
\ee
whatever $(a,b)$ taken for $\Xi(_{pi}^{qj})$, due to the fact that two successive fermion exchanges between two cobosons reduce to an identity.\

Equations (\ref{eq:H-NE0=xi},\ref{BnBmphiN}) give the Hamiltonian mean value $\lan H\ran_N$ as
\be
\frac{\lan \Phi_N| H|\Phi_N\ran}{\lan \Phi_N| \Phi_N\ran}\simeq NE_0+\frac{N(N-1)}{2}\frac{F_{N-2}}{F_N}\Big(\xi(_{00}^{ 00})-\xi^{in}(_{00}^{ 00})  \Big)\ \label{eq:HMF-NE0}
\ee
in agreement with previous work\cite{OBM<H>}. We also find from Eq.~(\ref{eq:H-NERN})
\be
\mathcal{E}_N\simeq \lan H\ran_N+
\frac{F_{N-2}}{F_N}
\sum_{ij\neq 00}\Big(\zeta (_{0i}^{ 0j}) +\Delta_N \lambda (_{0i}^{0j})\Big)c_{ij}\, \label{eq:mathEN=cij}
\ee
where, with the help of Eq.~(\ref{rel:xiinxiout}), the $\zeta(_{mi}^{\,nj})$ scattering is just the one defined in Eq.~(\ref{eq:zeta}). 

To obtain $c_{ij}$, we project Eq.~(\ref{eq:Scrhodingertwoexciton}) on $B_m^\dag B^\dag_n |\Phi_{N-2}\ran$. Using Eqs.~(\ref{eq:H-EijN-2}) and (\ref{eq:BBBB}), we find, after some algebra,
\bea
\lefteqn{(E_{00}-E_{mn}+\Delta_N)c_{mn}\simeq
 \frac{N(N-1)}{2}
\Big(  \zeta(_{m0}^{\, n0})\! +\!\Delta_N \lambda (_{m0}^{\,n0}) \Big)}\hspace{7.9cm}\nn\\
+\sum_{ij\neq00} \Big( \zeta(_{mi}^{\, nj})+\Delta_N \lambda (_{mi}^{\,nj}) \Big)c_{ij}\hspace{0.5cm}\label{eq:cmnsimeq}
\eea

Next, we drop the $\Delta_N $ terms: indeed, the change $\Delta_N $ in the energy of $N$ cobosons induced by interaction depends linearly on coboson number or equivalently on sample volume. So, the $\Delta_N $ terms provide contributions to $\mathcal{E}_N$ with improper volume dependence. They come from disconnected processes which cancel out, as standard in interaction expansion.

By setting $c_{ij}=d_{ij}N(N-1)/2$,  the above equation reduces to
  \be
  \hat\zeta(_{m0}^{\, n0})\equiv(E_{00}-E_{mn})d_{mn}\simeq
\zeta(_{m0}^{\, n0})\! +\sum_{ij\neq00}\zeta(_{mi}^{\, nj})d_{ij}\label{d_mn}\, ,
  \ee
As $F_{N-2}/F_N\simeq1$ at first order in density\cite{MoniqPhysreport}, this gives
 the $N$-coboson energy as
 \be
\mathcal{E}_N\simeq NE_0+ \frac{N(N-1)}{2}\zeta (_{00}^{00})\Big(1+\mathcal{F}\Big)\label{eq:EN-NE0F}
\ee
where $\mathcal{F}$ denotes the correction factor to the Born approximation
\be
\mathcal{F}= \frac{1}{\zeta (_{00}^{00})}\sum_{ij\neq 00}\zeta (_{0i}^{0j})d_{ij}=\frac{1}{\zeta (_{00}^{00})}\sum_{ij\neq 00}\frac{\zeta (_{0i}^{0j}) \hat \zeta (_{i0}^{j0})}{E_{00}-E_{ij}} \, \label{CorF}
\ee
Equation (\ref{eq:mathEN-NE0}) readily follows from the above equations.


In the following, we concentrate on Coulomb potential to address the ground-state energy of $N$ excitons made of same-spin electrons and same-spin holes. As previously shown\cite{MoniqPhysreport},
  the direct Coulomb scattering $ \xi(_{00}^{00})$ is equal to zero because repulsion between two electrons or between two holes is as strong as attraction between one electron and one hole. By contrast, the exchange Coulomb scattering $ \xi^{in}(_{00}^{ 00})$, equal to  $\xi^{out}(_{00}^{00})$ due to Eq.~(\ref{rel:xiinxiout}), is given by ${(a_B/L)}^D\xi_D$ in $3D$ Rydberg units, with $\xi_3=-26\pi/3$ and $\xi_2=-(8\pi-315\pi^3/512)$ for electrons and holes in the same quantum well\cite{MoniqueEL2007}. $\zeta(_{00}^{00})$ thus reduces in $3D$ to $-\xi^{in}(_{00}^{00})=(26\pi/3){(a_B/L)}^3$.

  This analytical result has already been obtained by Keldysh and Kozlov\cite{Keldysh1968} in the late 60's. One great advantage of the coboson many-body formalism over Bogoliubov-like procedure used up to now\cite{Comte1982}, is to provide a physical understanding to this term:  it comes from one exchange interaction inside the condensate with direct scattering reducing to zero in the case of Coulomb potential. To the best of our knowledge, explicit result for higher-order Coulomb processes has not been derived yet. According to Eq.~(\ref{eq:zeta}), these processes contain direct and exchange Coulomb scatterings as well as Pauli scattering multiplied by the excitation energy of the pair states at hand. Here also, the coboson many-body formalism allows catching the physics of these Coulomb processes readily.\

  To calculate contribution coming from more than one Coulomb interaction, we first note that the exciton pair $(i,j)$ excited from the condensate, which enters the repeated interaction of Eq.~(\ref{eq:zetaij00}), has a total center-of-mass momentum equal to zero, $\vQ_i+\vQ_j=\textbf{0}$, because Coulomb and exchange processes conserve momentum. These pairs can have arbitrary relative motion indices ($\nu_i,\nu_j$). However, due to energy denominators, dominant terms come from excitons staying in the relative motion ground state, $\nu_0$. So, the scatterings $\zeta(_{m\,i}^{n\,j})$ for the dominant processes in the ladder equation (\ref{eq:zetaij00}) reduce to $\zeta(_{\;\;\,\vQ_m,\nu_0\,\,\,\,\,\,\vQ_i,\nu_0}^{-\vQ_m,\nu_0\,\,-\vQ_i,\nu_0})\equiv\zeta( \vQ_m,\vQ_i)$.

  \begin{figure}[t!]
\begin{center}
   \includegraphics[trim=0cm 0.5cm 1cm 0.5cm,clip,width=3.6in] {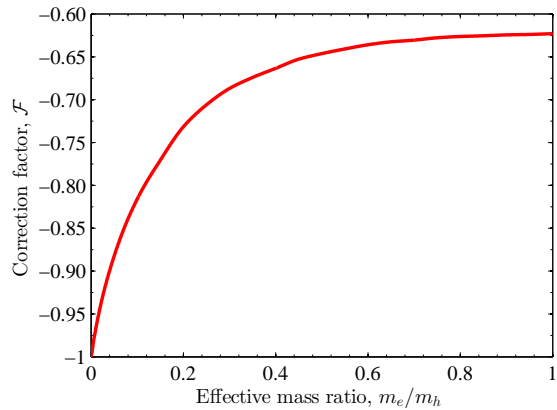}
   \caption{\small Correction factor $\mathcal{F}$ to the Born approximation as a function of electron-to-hole effective-mass ratio, $m_e/m_h$.}
   \label{Fig2}
\end{center}
\end{figure}

  We have numerically solved the integral equation (\ref{eq:zetaij00}) with $\zeta(_{m\,i}^{n\,j})$ replaced by $\zeta( \vQ_m,\vQ_i)$. This amounts to only keeping the $1$s ground state $\nu_0$ for the exciton relative motion while the center-of-mass momentum is treated using a dense mesh in {\bf Q}. Details of the numerical implementation is similar to what is described in Ref.~\onlinecite{SYAnnals}. Numerical results for the correction factor $\mathcal{F}$ to the Born approximation, defined in Eq.~(\ref{CorF}), are presented in Fig.~\ref{Fig2} as a function of electron-to-hole effective-mass ratio ($m_e/m_h$). It is seen that $\mathcal{F}$ monotonically decreases from  $\mathcal{F} \sim -0.62$ for $m_e/m_h=1$ down to $\mathcal{F}=-1$ when $m_e/m_h\rightarrow 0$, indicating a complete cancellation of the density-linear term in this limit.

  This cancellation remains true when all relative motion levels $\nu$ are kept in the $(i,j)$ sum of Eq.~(\ref{eq:zetaij00}). This can be shown analytically by considering Eq.~(\ref{d_mn}) for $m=({\bf Q},\nu_0)$ and $n=(-{\bf Q},\nu_0)$. The excited pair energy $E_{mn}-E_{00}$ goes to zero when the hole mass, \textit{i.e.}, the center-of-mass mass, goes to infinity. So, in this limit, the $(i,j)$ sum in Eq.~(\ref{d_mn}) must go to $-\zeta(_{\;\;\,\vQ,\nu_0\,0}^{-\vQ,\nu_0\,\,0})$ which, for $\vQ \rightarrow \textbf{0}$, reduces to $-\zeta(_{00}^{00})$. According to Eq.~(\ref{CorF}), $\mathcal{F}$ thus goes to $-1$ in the large hole mass limit. This physically means that, when the hole mass approaches infinity, the exciton gas does not resemble the $|\Psi\ran$ state given in Eq.~(\ref{psiNexpansion}) but should be better described as a 
    set of localized excitons. This limit will be studied elsewhere.


\textbf{To conclude}, we have derived the energy of $N$ composite bosons linear in density at all orders in fermion-fermion interaction, using the coboson many-body formalism which treats the Pauli exclusion principle in an exact way. This linear term follows from a ``ladder" integral equation that is numerically solved in the case of Coulomb potential between same-spin electrons and same-spin holes making semiconductor (triplet) excitons. We show that higher-order interactions drastically reduces the value of the density-linear term obtained within the Born approximation, down to zero in the infinite hole mass limit. This coboson many-body formalism opens the route to securely address the  far more complex next-order term in density expected to behave as $n^{3/2}$ instead of $n^{2}$ [see Ref.~\onlinecite{Leyronas2007}]. This formalism can also be used for other than Coulomb potential and for other fields than Semiconductor Physics, such as Cold Quantum Gases. \cite{Petrov2004,Leyronas2007,Alzetto2013}\

 M.C. wishes to acknowledge many fruitful visits to Academia Sinica in Taipei and to NCKU in Tainan. S.-Y.S. and Y.-C.C. have also benefited from various visits to INSP in Paris.	This work was supported in part by Ministry of Science and Technology, Taiwan under contract no.  NSC 101-2112-M-001-024-MY3.

\

\

\

\

\




\end{document}